\documentclass[12pt]{article}
\usepackage[sumlimits]{amsmath}
\usepackage{amsfonts,amssymb,amscd}
\usepackage[all]{xy}
\usepackage{graphicx}
\usepackage{color}

\newcommand{\ud}{\mathrm{d}}

 \numberwithin{equation}{section}

\begin{document}

\title{Dromion solutions of noncommutative Davey-Stewartson equations}
\author{Claire R. Gilson and Susan R. Macfarlane\\Department of Mathematics\\
University of Glasgow\\Glasgow G12 8QW\\UK} \date{}\maketitle
\abstract{We consider a noncommutative version of the
Davey-Stewartson equations and derive two families of
quasideterminant solution via Darboux and binary Darboux
transformations.  These solutions can be verified by direct
substitution.  We then calculate the dromion solutions of the
equations and obtain computer plots in a noncommutative setting.}
\section{Introduction}The Davey-Stewartson (DS) equations have become a
topic of much interest in recent years.  Derived by A. Davey and K.
Stewartson in $1974$ \cite{DS}, the system is nonlinear in
$(2+1)$-dimensions and describes the evolution of a
three-dimensional wave-packet on water of finite depth.  By carrying
out a suitable dimensional reduction, the system can be reduced to
the $(1+1)$-dimensional nonlinear Schr\"{o}dinger (NLS) equation.\\A
major development in the understanding of the DS equations came in
$1988$, when Boiti \textit{et al}. \cite{BLMP} discovered a class of
exponentially localised solutions (two-dimensional solitons) which
undergo a phase shift and possible amplitude change on interaction
with other solitons.  These were later termed \textit{dromions} by
Fokas and Santini \cite{FS}, derived from the Greek \textit{dromos}
meaning \textit{tracks}, to highlight that the dromions lie at the
intersection of perpendicular track-like plane waves.\\Multidromion
solutions to the DS system have been obtained using a variety of
approaches - the inverse scattering method \cite{FS}, Hirota's
direct method \cite{HH} and others. These solutions have been
determined both in terms of Wronskian \cite{HH} and Grammian
\cite{GN2} determinants.\\Additionally, there has been considerable
interest in noncommutative versions of the DS equations.  Hamanaka
\cite{Ham} derived a system with noncommutativity defined in terms
of the Moyal star product \cite{Moyal}, while more recently, Dimakis
and M\"{u}ller-Hoissen \cite{DMH3} determined a similar system from
a multicomponent KP hierarchy.  This then enabled calculation of
dromion solutions in the matrix case.\\
The strategy that we employ here, whereby we introduce
noncommutativity into an integrable nonlinear wave equation without
destroying the solvability, has previously been considered by others
in the field, for example by Lechtenfeld and Popov \cite{LP}, and by
Lechtenfeld, Popov \textit{et al} in \cite{LPetal}, where a
noncommutative version of the sine-Gordon
equation is discussed.\\
In this paper we are not concerned with the nature of the
noncommutativity, and derive a system of noncommutative DS equations
in the most general way by utilising the same Lax pair as in the
commutative case but assuming no commutativity of the dependent
variables.  This method has also been employed by Gilson and Nimmo
in \cite{GN} for the case of the noncommutative
Kadomtsev-Petviashvili (KP) equation.  We find that the
noncommutative DS system obtained in this manner corresponds to that
given in an earlier paper by Schultz, Ablowitz and Bar Yaacov
\cite{SAY}, where a quantum version of the DS equation is ultimately
discussed.\\We derive quasiwronskian and quasigrammian solutions of
this system in section \ref{sec:qdsoln} via Darboux and binary
Darboux transformations and, in section \ref{sec:verif}, verify
these solutions by direct substitution.\\We then use the
quasigrammian solution to determine a class of dromion solution and,
by specifying that certain parameters in the solution are of matrix
rather than scalar form, obtain dromion solutions in the
noncommutative case. We conclude with computer plots of these
dromion solutions.
\section{Noncommutative Davey-Stewartson equations}We consider
the system of commutative DS equations given by Ablowitz and Schultz
in \cite{ASc}, with Lax pair \begin{subequations}
\begin{align}\label{eq:L}L&=\partial_{x}-\varLambda+\sigma{J}\partial_{y},\\
\label{eq:M}M&=\partial_{t}-A+\frac{\textrm{i}}{\sigma}\varLambda\partial_{y}
-\textrm{i}J\partial_{yy},\end{align}\end{subequations}where
\begin{equation}J=\begin{pmatrix}1&0\\0&-1
\end{pmatrix},\, \varLambda=\begin{pmatrix}0&q(x,y,t)\\r(x,y,t)&0
\end{pmatrix}\end{equation}for $r=\pm{q}^{*}$ ($q^*$ denotes the complex conjugate of $q$)
and $A$ is a $2\times{2}$ matrix given by
\begin{equation}A=\begin{pmatrix}A_{1}&\displaystyle\frac{\textrm{i}}{2\sigma^{2}}(q_{x}-\sigma{q}_{y})\\
-\displaystyle\frac{\textrm{i}}{2\sigma^{2}}(r_{x}+\sigma{r}_{y})&A_{2}\end{pmatrix}.\end{equation}
We choose $\sigma=-1$ or $\sigma=\textrm{i}$ for the DSI and DSII
equations respectively.  By considering the same Lax pair
(\ref{eq:L},b) as is used in the commutative case and assuming no
commutativity of variables (we do not specify the nature of the
noncommutativity), we obtain the compatibility condition
\begin{subequations}
\begin{align}\label{eq:comm1}-A_{x}+[\varLambda,A]+\varLambda_{t}-\sigma{J}A_{y}+\frac{\textrm{i}}{\sigma}\varLambda\varLambda_{y}-\textrm{i}J\varLambda_{yy}&=0,\\
\label{eq:comm2}\frac{\textrm{i}}{\sigma}\varLambda_{x}+\sigma[A,J]-\textrm{i}J\varLambda_{y}&=0,\end{align}\end{subequations}from
which we generate a system of noncommutative Davey-Stewartson (ncDS)
eqations\begin{subequations}\label{eq:ds}
\begin{align}\label{eq:ds1}\textrm{i}q_{t}&=-\frac{1}{2\sigma^{2}}\left(q_{xx}+\sigma^{2}q_{yy}\right)+\textrm{i}(A_{1}q-qA_{2}),\\
\label{eq:ds2}\textrm{i}r_{t}&=\frac{1}{2\sigma^{2}}\left(r_{xx}+\sigma^{2}r_{yy}\right)-\textrm{i}(rA_{1}-A_{2}r),\\
\label{eq:A1}(\partial_{x}+\sigma\partial_{y})A_{1}&=-\frac{\textrm{i}}{2\sigma^{2}}(\partial_{x}-\sigma\partial_{y})(qr),\\
\label{eq:A2}(\partial_{x}-\sigma\partial_{y})A_{2}&=\frac{\textrm{i}}{2\sigma^{2}}(\partial_{x}+\sigma\partial_{y})(rq).\end{align}\end{subequations}
(Note that we obtain from the above system the nonlinear
Schr\"{o}dinger (NLS) equation \cite{AC}
\begin{equation}\textrm{i}{q_t+q_{yy}\pm{2}qrq=0}\end{equation}and its
corresponding complex conjugate by taking a dimensional reduction
$\partial_{x}=0$, with $A_{1}=\pm{\frac{\textrm{i}}{2}}qr,$
$A_{2}=\pm\frac{\textrm{i}}{2}rq$ \cite{Ham}).
\\
\
\\
\ For notational convenience, and to avoid the use of identities
later when verifying solutions, we introduce a $2\times{2}$ matrix
$S=(s_{ij})$ ($i,j=1,2$) such that $\varLambda=[J,\sigma{S}]$
\cite{Li}, and hence
\begin{equation}\label{eq:R1}S=\begin{pmatrix}s_{11}&\displaystyle\frac{q}{2\sigma}\\-\displaystyle\frac{r}{2\sigma}&s_{22}
\end{pmatrix}.\end{equation}Additionally, by setting
\begin{equation}\label{eq:A}A=\frac{\textrm{i}}{\sigma}S_{x}-\textrm{i}JS_{y},\end{equation} equation (\ref{eq:comm2}) is automatically
satisfied, and (\ref{eq:comm1}) becomes
\begin{multline}\label{eq:SJ}-\frac{\textrm{i}}{\sigma}S_{xx}+\textrm{i}JSS_{x}-\textrm{i}SJS_{x}-\textrm{i}S_{x}JS
+\textrm{i}S_{x}SJ+\textrm{i}\sigma{J}S_{y}JS\\
-\textrm{i}\sigma{J}S_{y}SJ+\sigma{J}S_{t}-\sigma{S}_{t}J
-\textrm{i}\sigma{J}SS_{y}J+\textrm{i}\sigma{S}JS_{y}J+\textrm{i}\sigma{J}S_{yy}J=0.\end{multline}
Note that this is essentially the noncommutative analogue of the
Hirota bilinear form (see for example \cite{H}) of the DS equations.
\section{Quasideterminants}Here we briefly recall some of the
properties of quasideterminants.  A more detailed analysis can be
found in the original papers \cite{GGRW,GR}.\\The notion of a
\textit{quasideterminant} was first introduced by Gelfand and Retakh
in \cite{GR} as a straightforward way to define the determinant of a
matrix with noncommutative entries. Many equivalent definitions of
quasideterminants exist, one such being a recursive definition
involving inverse minors. Let $A=(a_{ij})$ be an $n\times{n}$ matrix
with entries over a usually noncommutative ring $\mathcal{R}$. We
denote the $(i,j)^{\text{th}}$ quasideterminant by
$\mid\!{A}\!\mid_{ij}$, where
\begin{equation}\label{eq:Aij}\mid\!{A}\!\mid_{ij}\,=a_{ij}-r_{i}^{j}(A^{ij})^{-1}s_{j}^{i}.\end{equation}
Here, $A^{ij}$ is the $(n-1)\times(n-1)$ minor matrix obtained from
$A$ by deleting the $i^{\text{th}}$ row and the $j^{\text{th}}$
column (note that this matrix must be invertible),  $r_{i}^{j}$ is
the row vector obtained from the $i^{\text{th}}$ row of $A$ by
deleting the $j^{\text{th}}$ entry, and $s_{j}^{i}$ is the column
vector obtained from the $j^{\text{th}}$ column of $A$ by deleting
the $i^{\text{th}}$ entry.\\A common notation employed when
discussing quasideterminants is to `box' the expansion element, i.e.
we write
\begin{equation}\mid\!A\!\mid_{11}=\begin{vmatrix}\boxed{a_{11}} &
a_{12}\\a_{21} & a_{22}\end{vmatrix}\end{equation}to denote the
$(1,1)^{\textrm{th}}$ quasideterminant.  It should be noted that the
above expansion formula is also valid in the case of block matrices,
provided the matrix to be inverted is square; for example
considering a block matrix
\[\begin{pmatrix}N&B\\C&d\end{pmatrix},\]where $N$ is a square matrix over
$\mathcal{R}$, $B$ and $C$ are column and row vectors over
$\mathcal{R}$ of compatible lengths, and $d\in\mathcal{R}$, we
have\begin{equation}\begin{vmatrix}N&B\\C&\boxed{d}\end{vmatrix}=d-CN^{-1}B.\end{equation}
Quasideterminants also provide a useful formula for the inverse of a
matrix: for an invertible $n\times{n}$ matrix
$A=(a_{ij})\,(i,j=1,\ldots,n)$, the $(i,j)^{\textrm{th}}$ entry of
$A^{-1}$ is given by
\begin{equation}\label{eq:inv}(A^{-1})_{ij}=\bigl(\mid\!A\!\mid_{ji}\bigr)^{-1}.\end{equation}
When the elements of $A$ commute, the quasideterminant
$\mid\!A\!\mid_{ij}$ is not simply the determinant of $A$, but
rather a \textit{ratio} of determinants: it is well-known that, for
$A$ invertible, the $(j,i)^{\text{th}}$ entry of $A^{-1}$ is
\[(-1)^{i+j}\frac{\det{A^{ij}}}{\det{A}}.\]Then, by (\ref{eq:inv}),
we can easily see that
\begin{equation}\label{eq:comm}\mid\!A\!\mid_{ij}=(-1)^{i+j}\frac{\det{A}}{\det{A^{ij}}}\end{equation}
in the commutative case.
\section{Quasideterminant solutions via Darboux
transformations}\label{sec:qdsoln}
\subsection{Darboux transformations}
Here we give a brief overview of Darboux transformations.  Further
information can be found in, for example, \cite{MS}.\\We follow the
notation given in \cite{GN}.  Let $\theta_{1},\ldots,\theta_{n}$ be
a particular set of eigenfunctions of an operator $L$, and define
$\Theta=(\theta_{1},\ldots,\theta_{n})$ and
$\widehat\Theta=\Bigl(\theta_{j}^{(i-1)}\Bigr)$ for
$i,j=1,\ldots,n$, the $n\times{n}$ Wronskian matrix of
$\theta_{1},\ldots,\theta_{n}$, where $^{(k)}$ denotes the
$k^{\textrm{th}}$ $y$-derivative.\\To iterate the Darboux
transformation, let $\theta_{[1]}=\theta_{1}$ and $\phi_{[1]}=\phi$
be a general eigenfunction of $L_{[1]}=L$, with $L_{[1]}$ covariant
under the action of the Darboux transformation
$G_{\theta_{[1]}}=\partial_{y}-\theta_{[1]}^{(1)}\theta_{[1]}^{-1}$.
Then the general eigenfunctions $\phi_{[2]}$ for
$L_{[2]}=G_{\theta_{[1]}}L_{[1]}G_{\theta_{[1]}}^{-1}$ are given by
\begin{equation}\phi_{[2]}=G_{\theta_{[1]}}\bigl(\phi_{[1]}\bigr)=\phi_{[1]}^{(1)}-\theta_{[1]}^{(1)}\theta_{[1]}^{-1}\phi_{[1]}
,\end{equation}with
\begin{equation}\theta_{[2]}=\phi_{[2]}|_{\phi\rightarrow\theta_{2}.}\end{equation}Continuing
this process, after $n$ iterations ($n\geq{1}),$ the
$n^{\textrm{th}}$ Darboux transformation of $\phi$ is given by
\begin{equation}\phi_{[n+1]}=\phi_{[n]}^{(1)}-\theta_{[n]}^{(1)}\theta_{[n]}^{-1}\phi_{[n]},\end{equation}where
\begin{equation}\theta_{[k]}=\phi_{[k]}|_{\phi\rightarrow\theta_{k}}.\end{equation}
\subsection{Quasiwronskian
solution of ncDS using Darboux transformations}\label{subsec:qwsoln}
We now determine the effect of the Darboux transformation
$G_{\theta}=\partial_{y}-\theta_{y}\theta^{-1}$ on the Lax operator
$L$ given by (\ref{eq:L}), with $\theta$ an eigenfunction of $L$.
Corresponding results hold for the operator $M$ given by
(\ref{eq:M}).  $L$ is covariant with respect to the Darboux
transformation, and, by supposing that $L$ is transformed to a new
operator $\tilde{L}$, say, we calculate that the effect of the
Darboux transformation $\tilde{L}=G_{\theta}LG_{\theta}^{-1}$ is
such
that\begin{equation}\tilde{\varLambda}=\varLambda-\sigma[J,\theta_{y}\theta^{-1}].\end{equation}Recalling
that $\varLambda=[J,\sigma{S}]$, we have
$\tilde{S}=S-\theta_{y}\theta^{-1}$, and hence, after $n$ repeated
Darboux
transformations,\begin{equation}S_{[n+1]}=S_{[n]}-(\theta_{[n]})_{y}\theta_{[n]}^{-1},\end{equation}where
$S_{[1]}=S$, $\theta_{[1]}=\theta$.  We express $S_{[n+1]}$ in
quasideterminant form as
\begin{align}\label{eq:qd}S_{[n+1]}&=S+\begin{vmatrix}\theta_{1}&\ldots&\theta_{n}&\begin{array}
{cc}0&0\\0&0\end{array}\\\vdots&{}&\vdots&\vdots\\
\theta_{1}^{(n-2)}&\ldots&\theta_{n}^{(n-2)}&\begin{array}
{cc}0&0\\0&0\end{array}\\
\theta_{1}^{(n-1)}&\ldots&\theta_{n}^{(n-1)}&\begin{array}
{cc}1&0\\0&1\end{array}\\
\theta_{1}^{(n)}&\ldots&\theta_{n}^{(n)}&\boxed{\begin{array}
{cc}0&0\\0&0\end{array}}\end{vmatrix},\end{align}where $^{(k)}$
denotes the $k^{\textrm{th}}$ $y$-derivative.  It should be noted
here that each $\theta_{i}$ ($i=1,\ldots,n$) is not a single entry
but a $2\times{2}$ matrix (since the $\theta_{i}$ are eigenfunctions
of $L,M$).  The Wronskian-like quasideterminant in (\ref{eq:qd}) is
termed a \textit{quasiwronskian}, see \cite{GN}, and \cite{FN} for
details of Wronskian determinants.\\For ease of notation, for
integers $i,j=1,\ldots,n$, we denote by $Q(i,j)$ the
quasideterminant
\cite{GN}\begin{equation}\label{eq:Qij}Q(i,j)=\begin{vmatrix}\widehat{\Theta}&\begin{array}{cc}f_{j}&e_{j}\end{array}\\\Theta^{(n+i)}&
\boxed{\begin{array}{cc}0&0\\0&0\end{array}}\end{vmatrix},\end{equation}where,
as before, $\widehat{\Theta}=(\theta_{j}^{(i-1)})_{i,j=1,\ldots,n}$
is the $n\times{n}$ Wronskian matrix of
$\theta_{1},\ldots,\theta_{n}$ and $^{(k)}$ denotes the
$k^{\textrm{th}}$ $y$-derivative, $\Theta$ is the row vector
$\left(\theta_{1},\ldots,\theta_{n}\right)$ of length $n$, and
$f_{j}$ and $e_{j}$ are $2n\times{1}$ column vectors with a $1$ in
the $(2n-2j-1)^{\textrm{th}}$ and $(2n-2j)^{\textrm{th}}$ row
respectively and zeros elsewhere. Again each $\theta_{i}$ is a
$2\times{2}$ matrix. In this definition of $Q(i,j)$, we allow $i,j$
to take any integer values subject to the convention that if either
$2n-2j$ or $2n-2j-1$ lies outside the range $1,2,\ldots,2n$, then
$e_{j}=f_{j}=0$ and so $Q(i,j)=0$. Hence (\ref{eq:qd}) can be
written as
\begin{equation}\label{eq:qd2}S=S_{0}+Q(0,0),\end{equation}where $S_{0}$ is any given solution of
the ncDS equations.  Here we choose the vacuum solution $S_{0}=0$
for simplicity.\\It will be useful to express the quasiwronskian
solution (\ref{eq:qd2}) in terms of the variables $q$ and $r$, the
variables in which the ncDS equations (\ref{eq:ds1}-d) are
expressed. We have
\begin{equation}\label{eq:RQ}S=Q(0,0),\end{equation}which gives, by applying the quasideterminant
expansion formula (\ref{eq:Aij}) and expressing each $\theta_{i}$
($i=1,\ldots,n$) as an appropriate $2\times{2}$ matrix
\begin{equation}\label{eq:thetai}\theta_{i}=\begin{pmatrix}\phi_{2i-1}&\phi_{2i}\\\psi_{2i-1}&\psi_{2i}\end{pmatrix}\end{equation}for
$\phi=\phi(x,y,t),\,\psi=\psi(x,y,t)$, an expression for $S$ in
terms of quasiwronskians, namely
\begin{equation}\label{eq:Rqd}S=\begin{pmatrix}\begin{vmatrix}\widehat\Theta&f_{0}
\\\phi^{(n)}&\boxed{0}\end{vmatrix}&
\begin{vmatrix}\widehat\Theta&e_{0}
\\\phi^{(n)}&\boxed{0}\end{vmatrix}\\{}&{}\\\begin{vmatrix}\widehat\Theta&f_{0}
\\\psi^{(n)}&\boxed{0}\end{vmatrix}&\begin{vmatrix}\widehat\Theta&e_{0}
\\\psi^{(n)}&\boxed{0}\end{vmatrix}\end{pmatrix},\end{equation}where $\phi^{(n)}$, $\psi^{(n)}$
denote the row vectors
$\left(\phi_{1}^{(n)},\ldots,\phi_{2n}^{(n)}\right)$,
$\left(\psi_{1}^{(n)},\ldots,\psi_{2n}^{(n)}\right)$ respectively.
By comparing with (\ref{eq:R1}), we immediately see that $q,\,r$ can
be expressed as quasiwronskians, namely
\begin{align}\label{eq:qrqw}q&=2\sigma\begin{vmatrix}\widehat\Theta&e_{0}
\\\phi^{(n)}&\boxed{0}\end{vmatrix},&r&=-2\sigma\begin{vmatrix}\widehat\Theta&f_{0}
\\\psi^{(n)}&\boxed{0}\end{vmatrix}.\end{align}
\subsection{Binary Darboux transformations}\label{subsec:BDT}In order to define a
binary Darboux transformation, we consider the adjoint Lax pair of
the ncDS system (\ref{eq:ds1}-d).  The notion of adjoint can be
easily extended from the well-known matrix situation to any ring
$\mathcal{R}$.  An element $a\in\mathcal{R}$ has adjoint $a^{\dag}$,
where the adjoint has the following properties: if $\partial$ is a
derivative acting on $\mathcal{R}$, $\partial^{\dag}=-\partial$, and
for any product $AB$ of elements of, or operators on $\mathcal{R}$,
$(AB)^{\dag}=B^{\dag}A^{\dag}$. Thus the adjoint Lax pair for the
ncDS system is given by
\begin{subequations}
\begin{align}\label{eq:Ladj}L^{\dag}&=-\partial_{x}-\varLambda^{\dag}-\frac{1}{\sigma}J\partial_{y},\\
\label{eq:Madj}M^{\dag}&=-\partial_{t}-A^{\dag}+\textrm{i}\sigma(\varLambda_{y}^{\dag}+\varLambda^{\dag}\partial_{y})
+\textrm{i}J\partial_{yy}.\end{align}\end{subequations}We construct
a binary Darboux transformation in the usual manner (see for example
\cite{MS}) by introducing a potential $\Omega(\phi,\psi)$ satisfying
the relations
\begin{subequations}\label{eq:omega}\begin{align}\label{eq:omegay}\Omega(\phi,\psi)_{y}&=\psi^{\dag}\phi,\\\label{eq:omegax}\Omega(\phi,\psi)_{x}&=-\sigma\psi^{\dag}J\phi,\\
\label{eq:omegat}\Omega(\phi,\psi)_{t}&=\textrm{i}(\psi^{\dag}J\phi_{y}-\psi_{y}^{\dag}J\phi),\end{align}\end{subequations}
with $\phi$ an eigenfunction of $L,M$ and $\psi$ an eigenfunction of
$L^{\dag},M^{\dag}$.  This definition is also valid for non-scalar
eigenfunctions: if $\Phi$ is an $n$-vector and $\Psi$ an $m$-vector,
then $\Omega(\Phi,\Psi)$ is an $m\times{n}$ matrix.  We then define
a binary Darboux transformation $G_{\theta,\rho}$ by
\begin{equation}G_{\theta,\rho}=1-\theta\Omega(\theta,\rho)^{-1}\partial_{y}^{-1}\rho^{\dag}\end{equation}for
eigenfunctions $\theta$ of $L,M$ and $\rho$ of $L^{\dag},M^{\dag}$,
so that
\begin{equation}\phi_{[2]}=G_{\theta_{[1]},\rho_{[1]}}\bigl(\phi_{[1]}\bigr)=
\phi_{[1]}-\theta_{[1]}\Omega(\theta_{[1]},\rho_{[1]})^{-1}\Omega(\rho_{[1]},\phi_{[1]})
\end{equation}and
\begin{equation}\psi_{[2]}=G_{\theta_{[1]},\rho_{[1]}}\bigl(\psi_{[1]}\bigr)=\psi_{[1]}
-\rho_{[1]}\Omega(\theta_{[1]},\rho_{[1]})^{-\dag}\Omega(\theta_{[1]},\psi_{[1]})^{\dag}
,\end{equation}with
\begin{align}\theta_{[2]}&=\phi_{[2]}|_{\phi\rightarrow\theta_{2},}&\rho_{[2]}&=\psi_{[2]}|_{\psi\rightarrow\rho_{2}.}\end{align}After
$n\geq{1}$ iterations, the $n^{\textrm{th}}$ binary Darboux
transformation is given by
\begin{equation}\phi_{[n+1]}=
\phi_{[n]}-\theta_{[n]}\Omega(\theta_{[n]},\rho_{[n]})^{-1}\Omega(\rho_{[n]},\phi_{[n]})
\end{equation}and
\begin{equation}\psi_{[n+1]}=\psi_{[n]}
-\rho_{[n]}\Omega(\theta_{[n]},\rho_{[n]})^{-\dag}\Omega(\theta_{[n]},\psi_{[n]})^{\dag}
,\end{equation}with
\begin{align}\theta_{[n]}&=\phi_{[n]}|_{\phi\rightarrow\theta_{n}},&\rho_{[n]}&=\psi_{[n]}|_{\psi\rightarrow\rho_{n}.}
\end{align}
\subsection{Quasigrammian
solution of ncDS using binary Darboux
transformations}\label{subsec:qgsoln}We now determine the effect of
the binary Darboux transformation $G_{\theta,\rho}$ on the operator
$L$ given by (\ref{eq:L}), with $\theta$ an eigenfunction of $L$ and
$\rho$ an eigenfunction of $L^{\dag}$. Corresponding results hold
for the operator $M$ given by (\ref{eq:M}) and its corresponding
adjoint $M^{\dag}$.  The operator $L$ is transformed to a new
operator $\hat{L}$, say, where
\begin{equation}\hat{L}=G_{\theta,\rho}LG_{\theta,\rho}^{-1}.\end{equation}We
find that
$\hat{\varLambda}=\varLambda-\sigma[J,\theta_{y}\theta^{-1}]-\sigma[{\hat\theta}_{y}{\hat\theta}^{-1},J]$,
and hence, since $\varLambda=[J,\sigma{S}]$, it follows that
\begin{equation}\hat{S}=S-\theta\Omega(\theta,\rho)^{-1}\rho^{\dag}\end{equation}with
$\hat{\theta}=-\theta\Omega(\theta,\rho)^{-1}$.  After $n$ repeated
applications of the binary Darboux transformation $G_{\theta,\rho}$,
we obtain
\begin{equation}S_{[n+1]}=S_{[n]}-\sum_{k=1}^{n}\theta_{[k]}\Omega(\theta_{[k]},\rho_{[k]})^{-1}
\rho_{[k]}^{\dag},\end{equation}where $S_{[1]}=S$,
$\theta_{[1]}=\theta$.  Defining
$\Theta=(\theta_{1},\ldots,\theta_{n})$ and
$P=(\rho_{1},\ldots,\rho_{n})$, we express $S_{[n+1]}$ in
\textit{quasigrammian} form \cite{GN} as
\begin{equation}\label{eq:qg}S_{[n+1]}=S+\begin{vmatrix}\Omega(\Theta,P)&P^{\dag}\\\Theta&\boxed{\begin{array}
{cc}0&0\\0&0\end{array}}\end{vmatrix},\end{equation}where $\Omega$
is the Grammian-like matrix defined by (\ref{eq:omega}).  Note that,
for $i=1,\ldots,n,$ each $\theta_{i},\,\rho_{i}$ is a $2\times{2}$
matrix (since the $\theta_{i},\,\rho_{i}$ are eigenfunctions of
$L,M$ and $L^{\dag},M^{\dag}$ respectively).\\For integers
$i,j=1,\ldots,n$, denote by $R(i,j)$ the quasigrammian
\cite{GN}\begin{equation}\label{eq:Rij1}R(i,j)=(-1)^{j}\begin{vmatrix}\Omega(\Theta,P)
&P^{\dag\,(j)}\\\Theta^{(i)}&\boxed{\begin{array}{cc}0&0\\0&0\end{array}}\end{vmatrix},\end{equation}so
that, by once again choosing a trivial vacuum for simplicity,
(\ref{eq:qg}) can be expressed as
\begin{equation}\label{eq:qg2}S=R(0,0).\end{equation}As in the quasiwronskian case, we apply
the quasideterminant expansion formula (\ref{eq:Aij}), choosing the
matrices $\theta_{i}$ ($i=1,\ldots,n$) as in (\ref{eq:thetai}) and
$P=\Theta{H}^{\dag}$, where $H$ is a constant square matrix, in this
case $2n\times{2n}$, which we assume to be invertible, with
$H^{\dag}$ denoting the Hermitian conjugate of $H$. Thus
\begin{equation}\label{eq:Sqg}S=\begin{pmatrix}\begin{vmatrix}\Omega(\Theta,P)&H\phi^{\dag}
\\\phi&\boxed{0}\end{vmatrix}&
\begin{vmatrix}\Omega(\Theta,P)&H\psi^{\dag}
\\\phi&\boxed{0}\end{vmatrix}\\{}&{}\\\begin{vmatrix}
\Omega(\Theta,P)&H\phi^{\dag}
\\\psi&\boxed{0}\end{vmatrix}&\begin{vmatrix}\Omega(\Theta,P)&H\psi^{\dag}
\\\psi&\boxed{0}\end{vmatrix}\end{pmatrix},\end{equation}where
$\phi$, $\psi$ denote the row vectors
$\left(\phi_{1},\ldots,\phi_{2n}\right)$,
$\left(\psi_{1},\ldots,\psi_{2n}\right)$ respectively, which gives,
by comparing the above matrix with (\ref{eq:R1}), quasigrammian
expressions for $q$, $r$, namely
\begin{align}\label{eq:qrqg}q&=2\sigma\begin{vmatrix}\Omega(\Theta,P)&H\psi^{\dag}
\\\phi&\boxed{0}\end{vmatrix},&r&=-2\sigma\begin{vmatrix}\Omega(\Theta,P)&H\phi^{\dag}
\\\psi&\boxed{0}\end{vmatrix}.\end{align}Thus we have obtained, in
(\ref{eq:qrqw}), expressions for $q$, $r$ in terms of
quasiwronskians, and in (\ref{eq:qrqg}), expressions in terms of
quasigrammians.  We now show how these solutions can be verified by
direct substitution by firstly explaining the procedure used to
determine the derivative of a quasideterminant.
\section{Derivatives of a quasideterminant}\label{sec:derivqd}We consider a general quasideterminant of the form
\begin{equation}\label{eq:Xi}\Xi=\begin{vmatrix}A&B\\C&\boxed{D}\end{vmatrix},\end{equation}where
$A$, $B$, $C$ and $D$ are matrices of size $2n\times{2n}$,
$2n\times{2}$, $2\times{2n}$ and $2\times{2}$ respectively. If $A$
is a Grammian-like matrix with derivative
\begin{equation}A'=\sum_{i=1}^{k}E_{i}F_{i},\end{equation} where
$E_{i}$ $(F_{i})$ are column (row) vectors of comparable lengths, it
can be shown that (see Appendix)
\begin{equation}\label{eq:lambdadashgram}\Xi'=\begin{vmatrix}
A&B\\C'&\boxed{D'}\end{vmatrix}+\begin{vmatrix}
A&B'\\C&\boxed{0}\end{vmatrix}+\sum_{i=1}^{k}\begin{vmatrix}
A&E_{i}\\C&\boxed{0}\end{vmatrix}\begin{vmatrix}
A&B\\F_{i}&\boxed{0}\end{vmatrix},\end{equation}where $`0$'
denotes the $2\times{2}$ matrix $\bigl( \begin{smallmatrix} 0&0\\
0&0
\end{smallmatrix} \bigr)$.  If $A$ does not have a Grammian-like structure, we find that
\begin{equation}\label{eq:lambdadashwron}\Xi'=\begin{vmatrix}A&B\\C'&\boxed{D'}\end{vmatrix}+\sum_{k=0}^{n-1}
\begin{vmatrix}A&\begin{array}{cc}f_{k}&e_{k}\end{array}\\C&\boxed{\begin{array}{cc}0&0\\0&0\end{array}}\end{vmatrix}
.\begin{vmatrix}A&B\\\begin{array}{c}(A^{2n-2k-1})'\\(A^{2n-2k})'\end{array}&
\boxed{\begin{array}{c}(B^{2n-2k-1})'\\(B^{2n-2k})'\end{array}}\end{vmatrix},
\end{equation}where $A^{k}$ denotes the $k^{\textrm{th}}$ row of $A$.
The formulae (\ref{eq:lambdadashgram}) and (\ref{eq:lambdadashwron})
can be utilised to obtain expressions for the derivatives of the
quasideterminants $Q(i,j)$ (detailed in the Appendix), namely
\begin{subequations}\label{eq:Qyxt}\begin{align}Q(i,j)_{y}&=Q(i+1,j)-Q(i,j+1)+Q(i,0)Q(0,j),\\
Q(i,j)_{x}&=-\sigma\left({J}Q(i+1,j)-Q(i,j+1)J+Q(i,0)JQ(0,j)\right),\\
Q(i,j)_{t}&=\textrm{i}\left(JQ(i+2,j)-Q(i,j+2)J
+Q(i,1)JQ(0,j)+Q(i,0)JQ(1,j)\right).\end{align}\end{subequations} It
turns out that the derivatives of $R(i,j)$ match exactly with those
of $Q(i,j)$, hence subsequent calculations to verify the
quasiwronskian solution of the ncDS equations will also be valid in
the quasigrammian case, meaning that we need only verify one case.
\section{Direct verification of quasiwronskian and quasigrammian
solutions}\label{sec:verif}We now show that
\begin{equation}S=Q(0,0)\qquad\qquad\text{and}\qquad\qquad{S}=R(0,0)\end{equation}are solutions of
the ncDS system (\ref{eq:ds1}-d), where $S$ is the $2\times{2}$
matrix given by (\ref{eq:R1}) and $\varLambda=[J,\sigma{S}]$. Using
the derivatives of $Q(i,j)$ obtained in section \ref{sec:derivqd},
we have, on setting $i=j=0$,
\begin{subequations}\label{eq:Sderiv}\begin{align}{S}_{y}&=Q(0,0)_{y}=Q(1,0)-Q(0,1)+Q(0,0)^{2},\\
{S}_{x}&=Q(0,0)_{x}=-\sigma\left({J}Q(1,0)-Q(0,1)J+Q(0,0)JQ(0,0)\right),\\
{S}_{t}&=Q(0,0)_{t}=\textrm{i}\left(JQ(2,0)-Q(0,2)J+Q(0,1)JQ(0,0)+Q(0,0)JQ(1,0)\right),\\
\begin{split}{S}_{yy}&=Q(2,0)+Q(0,2)-2Q(1,1)-Q(0,1)Q(0,0)+Q(0,0)Q(1,0)\\
&+2\{Q(1,0)Q(0,0)-Q(0,0)Q(0,1)\}+2Q(0,0)^{3},\end{split}\\
\begin{split}{S}_{xx}&={\sigma}^2\bigl(Q(2,0)+Q(0,2)-2JQ(1,1)J-Q(0,1)Q(0,0)+Q(0,0)Q(1,0)\\
&-2\{Q(0,0)JQ(0,1)J-JQ(1,0)JQ(0,0)\}+2Q(0,0)JQ(0,0)JQ(0,0)\bigr).\end{split}
\end{align}\end{subequations}
Substituting the above in (\ref{eq:SJ}), all terms cancel exactly
and thus the quasiwronskian solution $S=Q(0,0)$ is verified.  As
mentioned previously, we obtain the same derivative formulae whether
we use the quasiwronskian or quasigrammian formulation, and hence
the above calculation also confirms the validity of the
quasigrammian solution $S=R(0,0)$.
\section{Dromion solutions}To obtain dromion solutions of the
system of ncDS equations, we use the quasigrammian solution
$S=R(0,0)$ rather than the quasiwronskian solution since
verification of reality conditions is simpler.
\subsection{$(n,n)$-dromion solution - noncommutative case}We modify
the approach of \cite{GN2}, where dromion solutions of a system of
commutative DS equations were determined.  We consider the ncDS
system (\ref{eq:ds1}-d) and, by specifying that certain parameters
in the quasigrammian are of matrix rather than scalar form, we are
able to obtain dromion solutions valid in the noncommutative case.
Due to the complexity of this solution compared to the scalar case
considered in \cite{GN2}, we look in some detail only at the
simplest cases of the $(1,1)$- and $(2,2)$-dromion solutions.  We do
however verify reality for the general case.  Note here that to
obtain dromion solutions,
we consider the DSI case, and hence choose $\sigma=-1$.\\
\
\\
\ Recall the expressions for $q$, $r$ obtained in terms of
quasigrammians in (\ref{eq:qrqg}), namely
\begin{align}\label{eq:qrqg2}q&=-2\begin{vmatrix}\Omega(\Theta,P)&H\psi^{\dag}
\\\phi&\boxed{0}\end{vmatrix},&r&=2\begin{vmatrix}\Omega(\Theta,P)&H\phi^{\dag}
\\\psi&\boxed{0}\end{vmatrix},\end{align}where
$\phi$, $\psi$ denote the row vectors
$\left(\phi_{1},\ldots,\phi_{2n}\right)$,
$\left(\psi_{1},\ldots,\psi_{2n}\right)$ respectively and
$H=(h_{ij})$ is a constant square invertible matrix, with $^\dag$
denoting conjugate transpose.  By once again considering the
dispersion relations for the system, we are able to choose
expressions for $\phi,$ $\psi$ corresponding to dromion solutions.
From (\ref{eq:Thetadisp}) and the definition of
$\Theta=(\theta_1,\ldots,\theta_{n})$, where $\theta_{i}$ is given
by (\ref{eq:thetai}), it follows that $\phi,$ $\psi$ satisfy the
relations
\begin{subequations}\label{eq:phipsirel}\begin{align}(\phi_{j})_{x}&=(\phi_{j})_{y},&(\phi_{j})_{t}&=\textrm{i}(\phi_{j})_{yy},\\
(\psi_{j})_{x}&=-(\psi_{j})_{y},&(\psi_{j})_{t}&=-\textrm{i}(\psi_{j})_{yy}.\end{align}\end{subequations}
(Since the dispersion relations for $P$ are the same as those for
$\Theta$ when $\sigma=-1$ (see (\ref{eq:Thetadisp}),
(\ref{eq:Pdisp})), considering $P$ rather than $\Theta$ and
recalling that $P=\Theta{H}^{\dag}$ will give the same relations
(\ref{eq:phipsirel})).\\
\ \\
\ So far we have not specified the nature of the noncommutativity we
are considering.  One of the most straightforward cases to consider
is to express our fields $q$ and $r$ as $2\times{2}$ matrices.  Thus
for dromion solutions in the noncommutative case, we choose
\begin{subequations}\label{eq:phipsimatrix}\begin{align}\phi_{j}&=\alpha_{j}I_{2},\\\psi_{j}&=\beta_{j}I_2,\end{align}
\end{subequations}where $I_{2}$ denotes the $2\times{2}$ identity
matrix, and $\alpha_{j}$, $\beta_{j}$ the
exponentials \cite{GN2,HH}\begin{subequations}\begin{align}\alpha_{j}&=\exp(p_{j}x+\textrm{i}p_{j}^{2}t+p_{j}y+\alpha_{j_{0}}),\\
\beta_{j}&=\exp(q_{j}x-\textrm{i}q_{j}^{2}t-q_{j}y+\beta_{j_{0}}),\end{align}\end{subequations}for
$j=1,\ldots,2n$, suitable phase constants $\alpha_{j_{0}}$,
$\beta_{j_{0}}$ and constants $p_{j}$, $q_{j}$, whose real parts are
taken to be positive in order to give the correct asymptotic
behaviour.  The matrix $H$ can be assumed to have unit diagonal
since we are free to choose the phase constants $\alpha_{j_{0}},$
$\beta_{j_{0}}$ arbitrarily \cite{GN2}.  Using the coordinate
transformation $X=x+y,$ $Y=-(x-y)$, we have
\begin{subequations}\label{eq:phipsiXY}\begin{align}\alpha_{j}&=\exp(p_{j}X+\textrm{i}p_{j}^{2}t+\alpha_{j_{0}}),\\
\beta_{j}&=\exp(-q_{j}Y-\textrm{i}q_{j}^{2}t+\beta_{j_{0}}),\end{align}\end{subequations}so
that $\alpha_{j}=\alpha_{j}(X,t)$, $\beta_{j}=\beta_{j}(Y,t)$.\\We
now choose to simplify our notation so that we are working with only
$\phi_{1},\ldots,\phi_{n}$ and $\psi_{1},\ldots,\psi_{n}$ by
relabeling $\phi_{j}$ as $\phi_{\frac{j+1}{2}}$ for odd $j$ (i.e.
$j=1,3,\ldots,2n-1$) and setting $\phi_{j}=0$ for even $j$
($j=0,2,\ldots,2n)$, and similarly relabeling $\psi_{j}$ as
$\psi_{\frac{j}{2}}$ for even $j$ and  setting $\psi_{j}=0$ for odd
$j$, so that $\theta_{j}=\text{diag}(\phi_{j},\psi_{j})$
($j=1,\ldots,n$) and
\begin{subequations}\begin{align}\phi&=\begin{pmatrix}\phi_{1}&0&\phi_{2}&0&\ldots&\phi_{n}&0\end{pmatrix},\\
\psi&=\begin{pmatrix}0&\psi_{1}&0&\psi_{2}&\ldots&0&\psi_{n}\end{pmatrix},\end{align}\end{subequations}
where each $\phi_{j}$, $\psi_{j}$ is a $2\times{2}$ matrix as
defined in (\ref{eq:phipsimatrix}) above. Thus, for $n=1$, $q$
(which we henceforth denote by $q^{1}$ for the $(1,1)$-dromion case
and $q^n$ for the $(n,n)$-dromion case) can be expressed in
quasigrammian form as
\begin{equation}q^1=-2\begin{vmatrix}\Omega(\Theta,P)&H\begin{array}{cc}0&0\\0&0\\\beta_{1}^{*}&0\\0&\beta_{1}^{*}\end{array}\\
\begin{array}{cccc}\alpha_{1}&0&0&0\\0&\alpha_{1}&0&0\end{array}&\boxed{\begin{array}{cc}0&0\\0&0\end{array}}\end{vmatrix},\end{equation}
where $H=(h_{ij})$ is a constant invertible $4\times{4}$ matrix.
Applying the quasideterminant expansion formula (\ref{eq:Aij})
allows us to express $q^1$ as a $2\times{2}$ matrix, where each
entry is a quasigrammian, namely
\begin{align}\notag{q^1}&=-2\begin{pmatrix}\begin{vmatrix}\Omega(\Theta,P)&\begin{array}{c}h_{13}\beta_{1}^{*}\\\vdots
\\h_{43}\beta_{1}^{*}\end{array}\\\begin{array}{cccc}\alpha_{1}&0&0&0\end{array}&\boxed{0}\end{vmatrix}&
\begin{vmatrix}\Omega(\Theta,P)&\begin{array}{c}h_{14}\beta_{1}^{*}\\\vdots
\\h_{44}\beta_{1}^{*}\end{array}\\\begin{array}{cccc}\alpha_{1}&0&0&0\end{array}&\boxed{0}\end{vmatrix}\\{}\\
\begin{vmatrix}\Omega(\Theta,P)&\begin{array}{c}h_{13}\beta_{1}^{*}\\\vdots
\\h_{43}\beta_{1}^{*}\end{array}\\\begin{array}{cccc}0&\alpha_{1}&0&0\end{array}&\boxed{0}\end{vmatrix}&
\begin{vmatrix}\Omega(\Theta,P)&\begin{array}{c}h_{14}\beta_{1}^{*}\\\vdots
\\h_{44}\beta_{1}^{*}\end{array}\\\begin{array}{cccc}0&\alpha_{1}&0&0\end{array}&\boxed{0}\end{vmatrix}\end{pmatrix}\\
\label{eq:qmatrix}&=-2\begin{pmatrix}q_{11}^1&q_{12}^1\\q_{21}^1&q_{22}^1\end{pmatrix},\text{
say}.\end{align}We consider each quasigrammian in turn, but as an
example we will look at the quasigrammian $q_{11}^1$.  We apply
(\ref{eq:comm}) to express $q_{11}^1$ as a ratio of determinants,
namely
\begin{equation}q_{11}^{1}=-2\frac{\begin{vmatrix}\Omega(\Theta,P)&\begin{array}{c}h_{13}\beta_{1}^{*}\\\vdots
\\h_{43}\beta_{1}^{*}\end{array}\\\begin{array}{cccc}\alpha_{1}&0&0&0\end{array}&0\end{vmatrix}}{\begin{vmatrix}
\Omega(\Theta,P)\end{vmatrix}}=-2\frac{G_{11}^{1}}{F},\text{
say}.\end{equation}(We have introduced the notation $G_{vw}^{n}$ and
$q_{vw}^{n}$ ($v,w=1,2$) to emphasise that we are considering the
$(v,w)^{\textrm{th}}$ entry of the expansion of $q^n$ in the
$(n,n)$-dromion case).  Although (\ref{eq:comm}) is valid only in
the commutative case, our assumption here is that the variables $q$,
$r$ in the system of DS equations, and also the parameters
$\phi_{j}$, $\psi_{j}$, are noncommutative.  The exponentials
$\alpha_{j}$, $\beta_{j}$ given by (\ref{eq:phipsiXY}) are clearly
commutative by definition, hence we are free to use the result
(\ref{eq:comm}).\\By expanding the quasigrammian $r$ in
(\ref{eq:qrqg2}) in a similar way and extracting the
$(1,1)^{\textrm{th}}$ entry, we obtain
\begin{equation}r_{11}^{1}=2\frac{\begin{vmatrix}\Omega(\Theta,P)&\begin{array}{c}h_{11}\alpha_{1}^{*}\\\vdots
\\h_{41}\alpha_{1}^{*}\end{array}\\\begin{array}{cccc}0&0&\beta_{1}&0\end{array}&0\end{vmatrix}}{\begin{vmatrix}
\Omega(\Theta,P)\end{vmatrix}}=2\frac{K_{11}^{1}}{F},\text{
say},\end{equation}with similar results for $q_{12}^{1}$,
$r_{12}^{1}$ etc.
\subsection{Reality conditions}\label{subsec:reality}In the $(n,n)$-dromion case, we have
\begin{subequations}\begin{align}q_{vw}^{n}&=-2\frac{G_{vw}^{n}}{F},\\
r_{vw}^{n}&=2\frac{K_{vw}^{n}}{F},\end{align}\end{subequations}for
$v,w=1,2$.  To verify reality, we must check that
$r_{vw}^{n}=\pm({q_{vw}^{n}})^{*}$, i.e. that $F$ is real, and
$(G_{vw}^{n})^{*}=\pm{K_{vw}^{n}}$, where $(G_{vw}^{n})^{*}$ denotes
the complex conjugate of $G_{vw}^{n}$.\\We do not give details here,
however we can show that $F$ can be expressed in the form
\begin{equation}F=\mid\!{I_{4n}}+H\Phi\!\mid,\end{equation}where $H$ is $4n\times{4n}$, constant and
invertible, and $\Phi$ is the $4n\times{4n}$ matrix
\begin{equation}\Phi=
\begin{pmatrix}\int_{-\infty}^{X}{\phi_{1}^{*}\phi_{1}\,\ud X}&0&\ldots&\int_{-\infty}^{X}{\phi_{1}^{*}\phi_{n}\,\ud X}&0\\
0&\int_{Y}^{\infty}{\psi_{1}^{*}\psi_{1}\,\ud Y}&\ldots&0&\int_{Y}^{\infty}{\psi_{1}^{*}\psi_{n}\,\ud Y}\\\vdots&\vdots&{}&\vdots&\vdots\\
\int_{-\infty}^{X}{\phi_{n}^{*}\phi_{1}\,\ud X}&0&\ldots&\int_{-\infty}^{X}{\phi_{n}^{*}\phi_{n}\,\ud X}&0\\
0&\int_{Y}^{\infty}{\psi_{n}^{*}\psi_{1}\,\ud
Y}&\ldots&0&\int_{Y}^{\infty}{\psi_{n}^{*}\psi_{n}\,\ud
Y}\end{pmatrix},\end{equation}remembering that each $\phi_{i}$,
$\psi_{i}$ ($i=1,\ldots,n$) is a $2\times{2}$ matrix and `$0$'
denotes the $2\times{2}$ matrix
$\left(\begin{smallmatrix}0&0\\0&0\end{smallmatrix}\right)$.  With
$F$ in this form, it is straightforward to show that $F$ is real so
long as $H$ is a Hermitian matrix (see \cite{Ratter}).  This
condition is also required for $(G_{vw}^{n})^{*}=\pm{K_{vw}^{n}}$ -
in fact, we find that $r_{vw}^{n}=-(q_{vw}^{n})^{*}$ so long as $H$
is Hermitian.  This agrees with the work of Gilson and Nimmo in
\cite{GN2}.
\subsection{$(1,1)$-dromion solution - matrix case}We now show
computer plots of the $(1,1)$-dromion solution in the noncommutative
case, where we choose $\phi_1$, $\psi_1$ to be $2\times{2}$ matrices
as in (\ref{eq:phipsimatrix}).  A suitable choice of the parameters
$p_1$, $q_1$ and of the $4\times{4}$ Hermitian matrix $H$ allows us
to obtain plots of the four quasigrammian solutions $q_{11}^1$,
$q_{12}^1$, $q_{21}^1$, $q_{22}^1$ as detailed in
(\ref{eq:qmatrix}).\\
We are restricted in our choice of $p_1$, $q_1$ and $H$ in that we
require $F\neq{0}$.  In particular, we derive conditions so that
$F>0$.  The determinant $F$ can be expanded in terms of minor
matrices of $H=(h_{ij})$ $(i,j=1,\ldots,4)$,
giving\begin{multline}F=1+P_{1}\left(h_{234}^{234}+h_{134}^{134}\right){e^{2\eta}}+Q_{1}\left(h_{124}^{124}+h_{123}^{123}\right){e^{-2\xi}}
+P_{1}^2h_{34}^{34}e^{4\eta}
\\+Q_{1}^{2}h_{12}^{12}e^{-4\xi}+P_{1}Q_{1}\left(h_{24}^{24}+h_{14}^{14}+h_{23}^{23}+h_{13}^{13}\right)e^{2\eta-2\xi}
\\+P_{1}^{2}Q_{1}\left(h_{4}^{4}+h_{3}^{3}\right)e^{4\eta-2\xi}+P_{1}Q_{1}^{2}\left(h_{2}^{2}+h_{1}^{1}\right)
e^{2\eta-4\xi}+P_{1}^{2}Q_{1}^{2}he^{4\eta-4\xi},\end{multline}where
$P_{1}=1/2\mathfrak{R}(p_1)$, $Q_{1}=1/2\mathfrak{R}(q_1)$,
$\eta=\mathfrak{R}\left(p_{1})(X-2\mathfrak{I}(p_1)t\right)$,
$\xi=\mathfrak{R}\left(q_{1})(Y-2\mathfrak{I}(q_1)t\right)$, and
$h_{ij\ldots}^{rs\ldots}$ denotes the minor matrix obtained by
removing rows $i,j,\ldots$ and columns $r,s,\ldots$ of $H$, where
$i,j,\ldots$, $r,s,\ldots\in\{1,2,3,4\}.$  We have used `$h$' to
indicate that no rows or columns have been removed, i.e.
$h=\det{H}.$  We can also obtain expressions for each $G_{vw}^{1}$
($v,w=1,2$), for instance
\begin{equation}G_{11}^{1}=\alpha_{1}\beta_{1}^{*}\left(h_{234}^{124}-P_{1}h_{34}^{14}e^{2\eta}
-Q_{1}h_{23}^{12}e^{-2\xi}-P_{1}Q_{1}h_{3}^{1}e^{2\eta-2\xi}\right),\end{equation}with
$\alpha_{1},\beta_{1}$ defined as in (\ref{eq:phipsiXY}).  Similar
expansions can be obtained for $G_{12}^{1},G_{21}^{1}$ and
$G_{22}^{1}$.  Thus, it can be seen that for $F>0$, we require
$\mathfrak{R}(p_1),$ $\mathfrak{R}(q_1)$ and each of the minor
matrices in the expansion of $F$ to be greater than zero. Suitable
choices of $p_{1}, q_{1}$ and $H$ have been used to obtain the
dromion plots shown later.  In addition to the matrix-valued fields
$q$ and $r$, there are also matrix-valued fields $A_{1}$ and $A_{2}$
in the ncDS system (\ref{eq:ds1}-d).  Plotting the derivatives of
these fields gives plane waves as follows.\\From (\ref{eq:Sqg}), we
have an expression for the $2\times{2}$ matrix $S$ in terms of
quasigrammians.  In the DSI case
$A=-\textrm{i}S_{x}-\textrm{i}JS_{y}$ by (\ref{eq:A}), therefore
substituting for $S$ using (\ref{eq:Sqg}) and equating matrix
entries gives quasigrammian expressions for $A_1$, $A_2$,
namely\begin{subequations}\begin{align}A_{1}&=-\textrm{i}\begin{vmatrix}\Omega&H\phi^{\dag}\\\phi&\boxed{0}\end{vmatrix}_{x}
-\textrm{i}\begin{vmatrix}\Omega&H\phi^{\dag}\\\phi&\boxed{0}\end{vmatrix}_{y},\\
A_{2}&=-\textrm{i}\begin{vmatrix}\Omega&H\psi^{\dag}\\\psi&\boxed{0}\end{vmatrix}_{x}
+\textrm{i}\begin{vmatrix}\Omega&H\psi^{\dag}\\\psi&\boxed{0}\end{vmatrix}_{y}.\end{align}\end{subequations}
We choose $\phi_{1}$, $\psi_{1}$ to be $2\times{2}$ matrices as
before so that the boxed expansion element `$0$' in each
quasigrammian is the $2\times{2}$ matrix
$\left(\begin{smallmatrix}0&0\\0&0\end{smallmatrix}\right)$.  Thus,
expanding each quasigrammian in the usual manner gives a
$2\times{2}$ matrix where each entry is a quasigrammian, and hence
we have distinct expressions for $A_{1}$, $A_{2}$ corresponding to
each of the four dromions $q_{11}^1,q_{12}^1,q_{21}^1,q_{22}^1$.
Considering (\ref{eq:A1},\,d), we find that plotting the combination
$(\partial_{x}+\sigma{\partial_{y}})A_1$ for $\sigma=-1$ gives a
plane wave travelling in the $X$ direction, while the combination
$(\partial_{x}-\sigma\partial_{y})A_2$ gives a plane wave in the $Y$
direction.  These, along with the dromions corresponding to each
plane wave, have been plotted at time $t=0$ in figures $1$ and $2$.
\begin{figure}[h]\begin{center}\includegraphics[width=5.3in,height=3.3in]{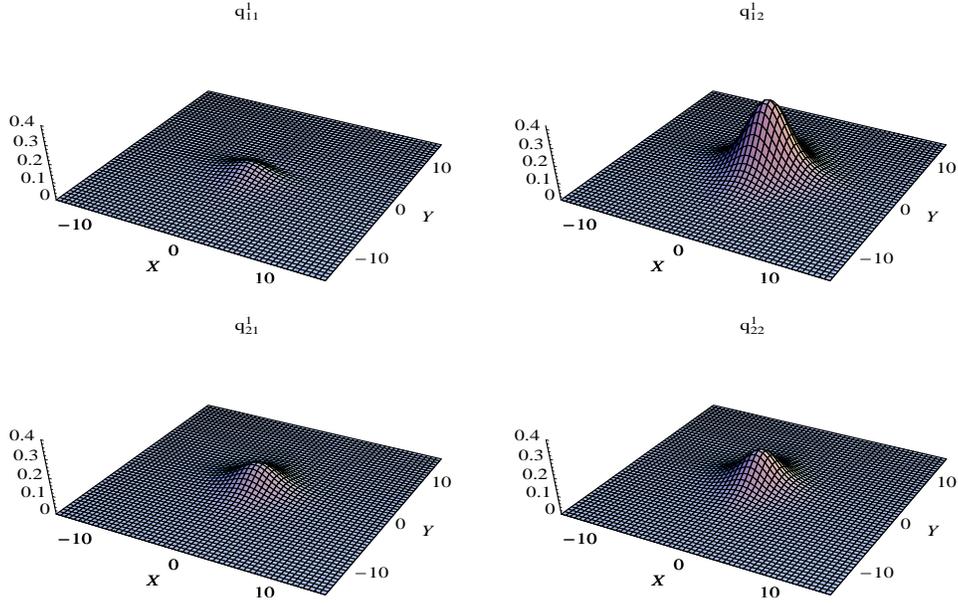}\caption{$(1,1)$-dromion plots with $p_1=\frac{1}{2}+\textrm{i}$, $q_1
=\frac{1}{2}-\textrm{i}$ and $h_{12}=\frac{1}{2}$,
$h_{13}=\frac{1}{4}$, $h_{14}=\frac{3}{4}$, $h_{23}=\frac{1}{3}$,
$h_{24}=\frac{1}{2}$, $h_{34}=\frac{1}{3}$.}\end{center}\end{figure}
\begin{figure}[h!]\begin{center}\includegraphics[width=5.3in,height=3.3in]{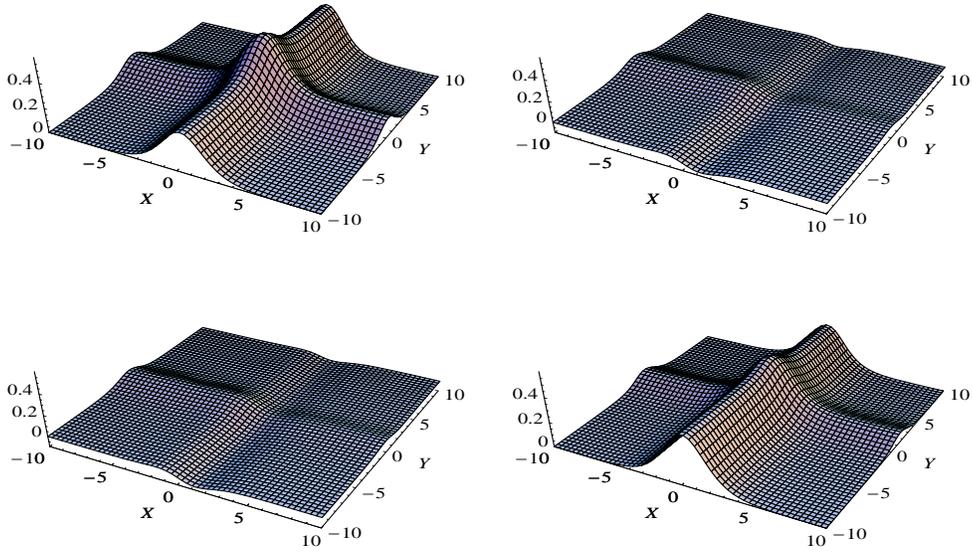}\caption{Plane waves
corresponding to, clockwise from top left, $q_{11}^1$, $q_{12}^1$,
$q_{22}^1$, $q_{21}^1$.}\end{center}\end{figure}\\
As can be seen from figure $1$, single dromions of differing heights
occur in each of the fields $q_{11}$, $q_{12}$, $q_{21}$ and
$q_{22}$. If we were to plot the $(1,1)$-dromion solution in the
commutative (scalar) case (that is, if we were to choose $q$ and its
complex conjugate to be of scalar rather than matrix form), we would
obtain only one dromion in the single field $q$. This dromion and
its plane waves would have the same basic structure as those above,
and thus there would be no marked difference in the appearance of
the dromions in the commutative and noncommutative cases. The main
difference between the two situations concerns the number of
parameters - a far greater number in the noncommutative case gives
us more freedom to control the heights of the dromions, however some
extra care has to be taken in choosing the parameters so that no
singularities occur in the solution.
 \subsection{$(2,2)$-dromion solution - matrix
case}\label{subsec:22dromion} In the scalar case \cite{GN2}, Gilson
and Nimmo carried out a detailed asymptotic analysis of their
$(M,N)$-dromion solution, and were able to obtain compact
expressions for the phase-shifts and changes in amplitude that occur
due to dromion interactions. They then used the results of this
analysis to study a class of $(2,2)$-dromions with scattering-type
interaction properties.  The Hermitian matrix $H$ could be chosen in
such a way so that some of the dromions had zero amplitude either as
$t\rightarrow{-\infty}$ or
as $t\rightarrow{+\infty}$.\\
\
\\
\ For the $(2,2)$-dromion solution in the matrix case, detailed
calculations of this type are more complicated due to the large
number of terms involved.  However, we can adopt the same approach
to carry out some of the more straightforward calculations.  In
particular, we obtain plots of the situation in which the
$(1,1)^{\textrm{th}}$ dromion in each of the solutions $q_{11}^{2}$,
$q_{12}^{2}$, $q_{21}^{2}$ and $q_{22}^{2}$ does not appear as
$t\rightarrow{-\infty}$.  These are depicted in figures $3$-$5$.\\
\
\\
\ To analyse this situation, we focus our attention on $q_{11}^{2}$
and consider $G_{11}^{2}$ in a frame moving with the
$(1,1)^{\textrm{th}}$ dromion.  We define
\begin{subequations}\label{eq:XhatYhat}\begin{align}\hat{X}&=X-2\Im({p_{1}})t,\\
\hat{Y}&=Y-2\Im({q_{1}})t,\end{align}\end{subequations}and consider
the limits of $G_{11}^{2}$ as $t\rightarrow{-\infty}$.  Let
\begin{subequations}\label{eq:eta2xi2}\begin{align}\notag\eta_{2}&=\Re(p_{2})\left(X-2\Im(p_{2})t\right)\\
&=\Re(p_{2})\left(\hat{X}-2(\Im(p_{2})-\Im(p_{1}))t\right),\intertext{and
similarly}\notag\xi_{2}&=\Re(q_{2})\left(Y-2\Im(q_{2})t\right)\\
&=\Re(q_{2})\left(\hat{Y}-2(\Im(q_{2})-\Im(q_{1}))t\right).
\end{align}\end{subequations}We choose to order the $p_{i},q_{i}$
($i=1,2$) by means of their imaginary parts, so that
$\Im(p_{1})>\Im(p_{2})$ and $\Im(q_{1})<\Im(q_{2})$.  Thus, as
$t\rightarrow{-\infty}$, $\eta_{2}\rightarrow{-\infty}$ and
$\xi_{2}\rightarrow{+\infty}$.  It can easily be shown that
$\eta_{2}$, $-\xi_{2}$ determine the real parts of the exponents in
$\alpha_{2}$, $\beta_{2}$ respectively, where $\alpha_{2}$,
$\beta_{2}$ are defined as in (\ref{eq:phipsiXY}), so that, as
$t\rightarrow{-\infty}$, $\alpha_{2}$, $\beta_{2}\rightarrow{0}$
(and hence $\alpha_{2}^{*}$, $\beta_{2}^{*}\rightarrow{0}$ also).
Therefore, by setting
$\alpha_{2},\alpha_{2}^{*},\beta_{2},\beta_{2}^{*}\rightarrow{0}$ in
$G_{11}^{2}$ and expanding the resulting determinant, we obtain a
compact expression for $G_{11}^{2}$ as $t\rightarrow{-\infty}$,
namely
\begin{equation}\label{eq:G112}G_{11}^{2}=-\alpha_{1}\beta_{1}^{*}\left(h_{2345678}^{1245678}-P_{1}h_{345678}^{145678}e^{2\eta}
-Q_{1}h_{235678}^{125678}e^{-2\xi}+P_{1}Q_{1}h_{35678}^{15678}e^{2\eta-2\xi}\right).\end{equation}
Similar expressions can be obtained for the other three determinants
$G_{12}^{2}$, $G_{21}^{2}$ and $G_{22}^{2}$ by considering an
extension of (\ref{eq:qmatrix}) to the $(2,2)$-dromion case and
interchanging columns appropriately: for example, we interchange
columns $3$ and $4$, and $7$ and $8$, in the expansion of
$G_{11}^{2}$ as $t\rightarrow{-\infty}$ to obtain an analogous
expression for $G_{12}^{2}$.  Thus we have, as
$t\rightarrow{-}\infty$, compact expressions for the minors of $H$
governing the $(1,1)^\textrm{th}$ dromion in each of $G_{11}^{2}$,
$G_{12}^{2}$, $G_{21}^{2}$, $G_{22}^{2}$,
namely\begin{multline}\begin{pmatrix}G_{11}^{2}&G_{12}^{2}\\
G_{21}^{2}&G_{22}^{2}\\\end{pmatrix}=-\alpha_{1}\beta_{1}^{*}\Biggl\{\begin{pmatrix}|h_{13}|&
|h_{14}|\\|h_{23}|&|h_{24}|
\end{pmatrix}+P_{1}T\begin{pmatrix}{\begin{vmatrix}h_{12}&h_{13}\\h_{22}&h_{23}\end{vmatrix}}&
{\begin{vmatrix}h_{12}&h_{14}\\h_{22}&h_{24}\end{vmatrix}}\\{}&{}\\
{\begin{vmatrix}h_{11}&h_{13}\\h_{21}&h_{23}\end{vmatrix}}&
{\begin{vmatrix}h_{11}&h_{14}\\h_{21}&h_{24}\end{vmatrix}}\end{pmatrix}e^{2\eta}\\+
Q_{1}T\begin{pmatrix}{\begin{vmatrix}h_{13}&h_{14}\\h_{43}&h_{44}\end{vmatrix}}&
{\begin{vmatrix}h_{13}&h_{14}\\h_{33}&h_{34}\end{vmatrix}}\\{}\\
{\begin{vmatrix}h_{23}&h_{24}\\h_{43}&h_{44}\end{vmatrix}}&
{\begin{vmatrix}h_{23}&h_{24}\\h_{33}&h_{34}\end{vmatrix}}\end{pmatrix}e^{-2\xi}+
P_{1}Q_{1}T\begin{pmatrix}{\begin{vmatrix}h_{12}&h_{13}&h_{14}\\h_{22}&h_{23}&h_{24}\\h_{42}&
h_{43}&h_{44}\end{vmatrix}}&
{\begin{vmatrix}h_{12}&h_{13}&h_{14}\\h_{22}&h_{23}&h_{24}\\h_{32}&
h_{33}&h_{34}\end{vmatrix}}\\{}\\
{\begin{vmatrix}h_{11}&h_{13}&h_{14}\\h_{21}&h_{23}&h_{24}\\h_{41}&
h_{43}&h_{44}\end{vmatrix}}&
{\begin{vmatrix}h_{11}&h_{13}&h_{14}\\h_{21}&h_{23}&h_{24}\\h_{31}&
h_{33}&h_{34}\end{vmatrix}}\end{pmatrix}Te^{2\eta-2\xi}\Biggr\},
\end{multline}where $T=\text{diag}(-1,1)$.  Since we have written out each minor matrix
explicitly, rather than using the abbreviated notation as in
(\ref{eq:G112}) above, it can easily be seen that, by setting each
of $h_{13},h_{14},h_{23},h_{24}$ equal to zero, the
$(1,1)^{\textrm{th}}$ dromion in each of
$q_{11}^{2},q_{12}^{2},q_{21}^{2}$ and $q_{22}^{2}$ will vanish as
$t\rightarrow{-\infty}$.  This is shown in figures $3$-$5$ below,
where we have chosen $p_{1}=q_{2}=\frac{1}{2}+\textrm{i}$,
$p_{2}=q_{1}=\frac{1}{2}-\textrm{i}$, and appropriate values of
$h_{ij}$ ($i,j=1,\ldots,8$).  (Note that in the plots of
$q_{11}^{2}$ and $q_{12}^{2}$ in figure $3$, two of the dromions
have very small amplitude).  We have also shown, in figure $6$, a
close-up of one of the dromion interactions at $t=0$.
\begin{figure}[h!]\begin{center}\includegraphics[width=5.3in,height=3.3in]{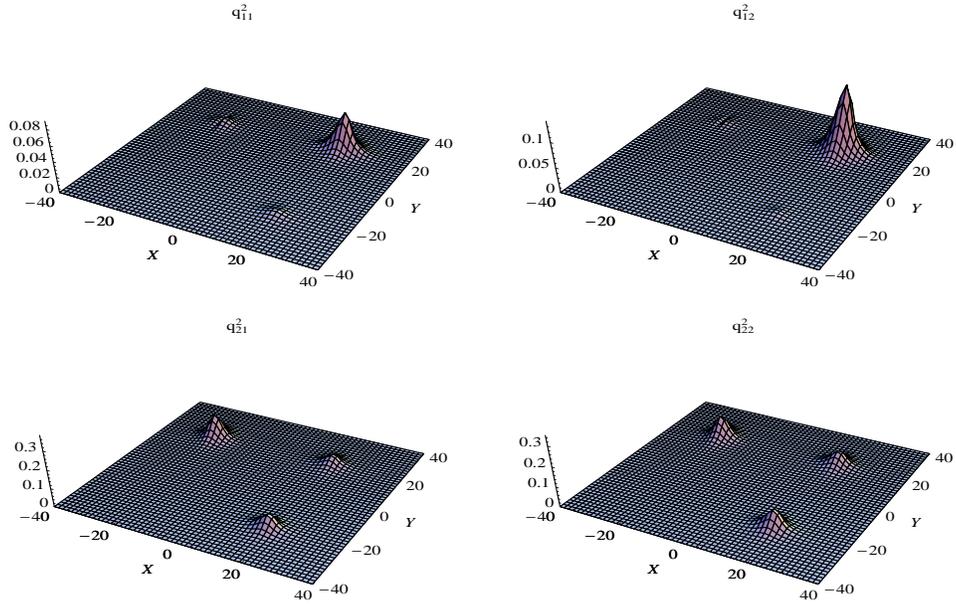}\caption{
$(2,2)$-dromion plots at $t=-10$.}\end{center}\end{figure}
\begin{figure}[h!]\begin{center}\includegraphics[width=5.3in,height=3.3in]{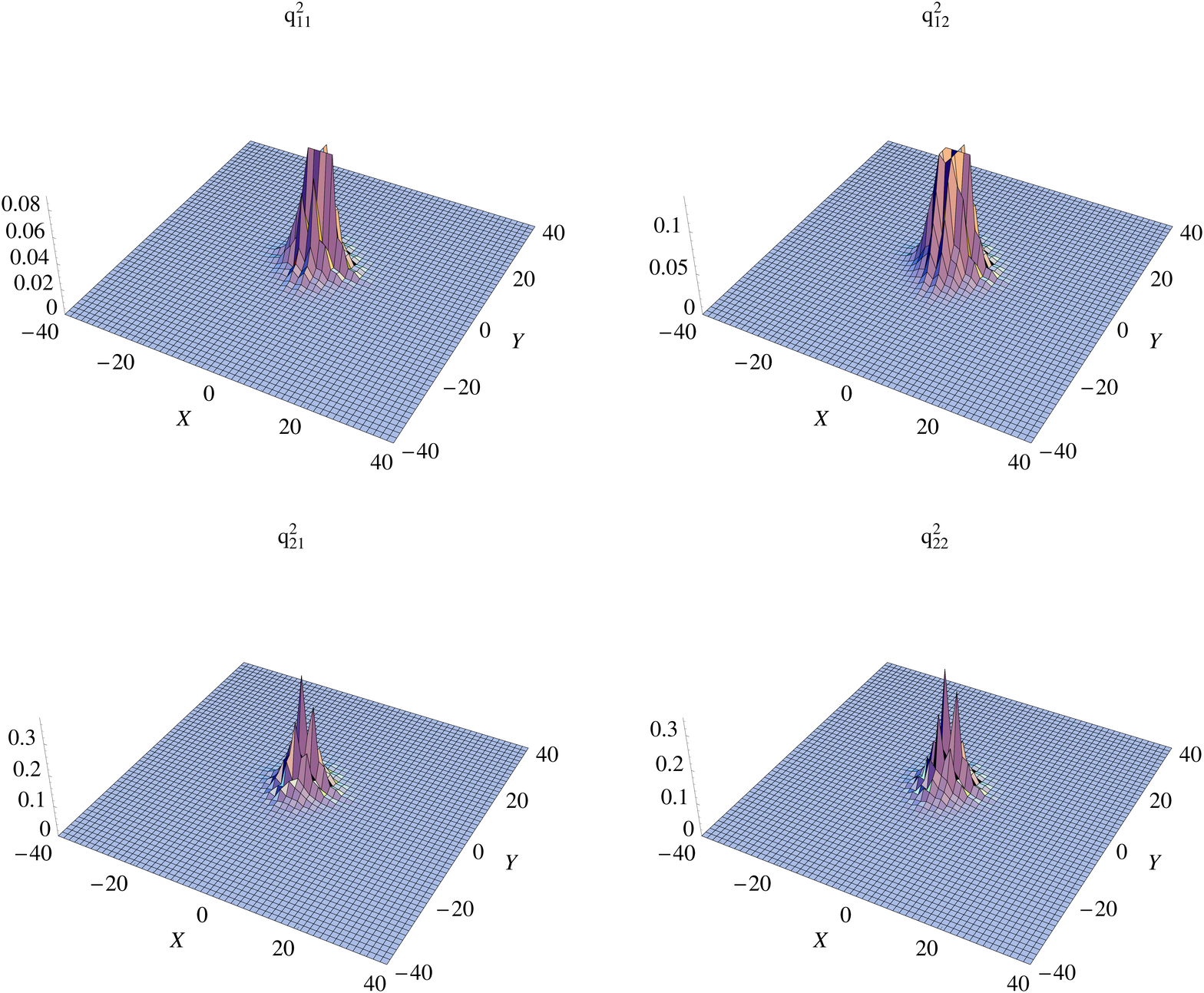}\caption{$(2,2)$-dromion plots at $t=0$.}\end{center}\end{figure}
\begin{figure}[h!]\begin{center}\includegraphics[width=5.3in,height=3.3in]{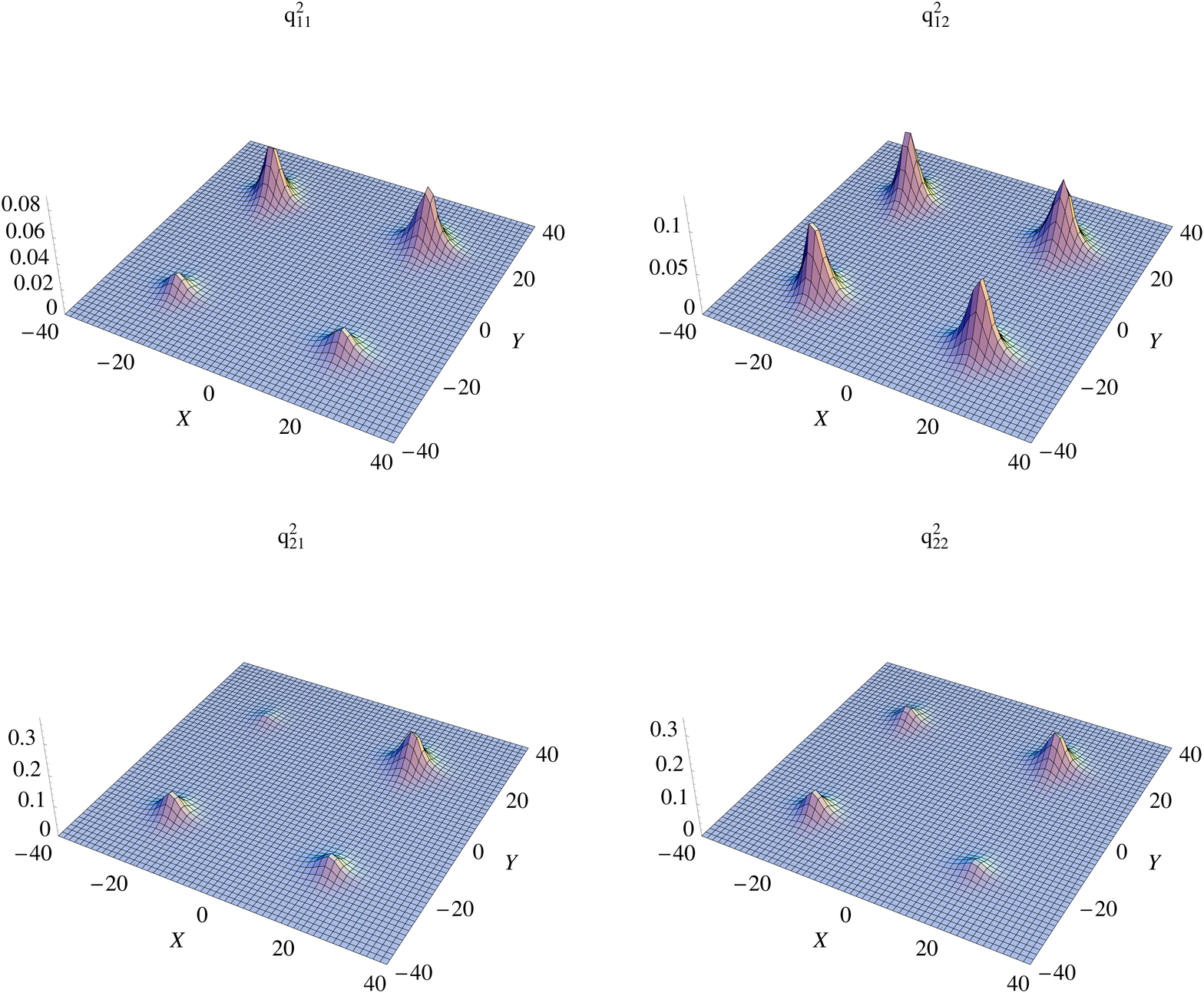}\caption{$(2,2)$-dromion plots at $t=10$.}\end{center}\end{figure}
\begin{figure}[h!]\begin{center}\includegraphics[width=3.3in,height=2.3in]{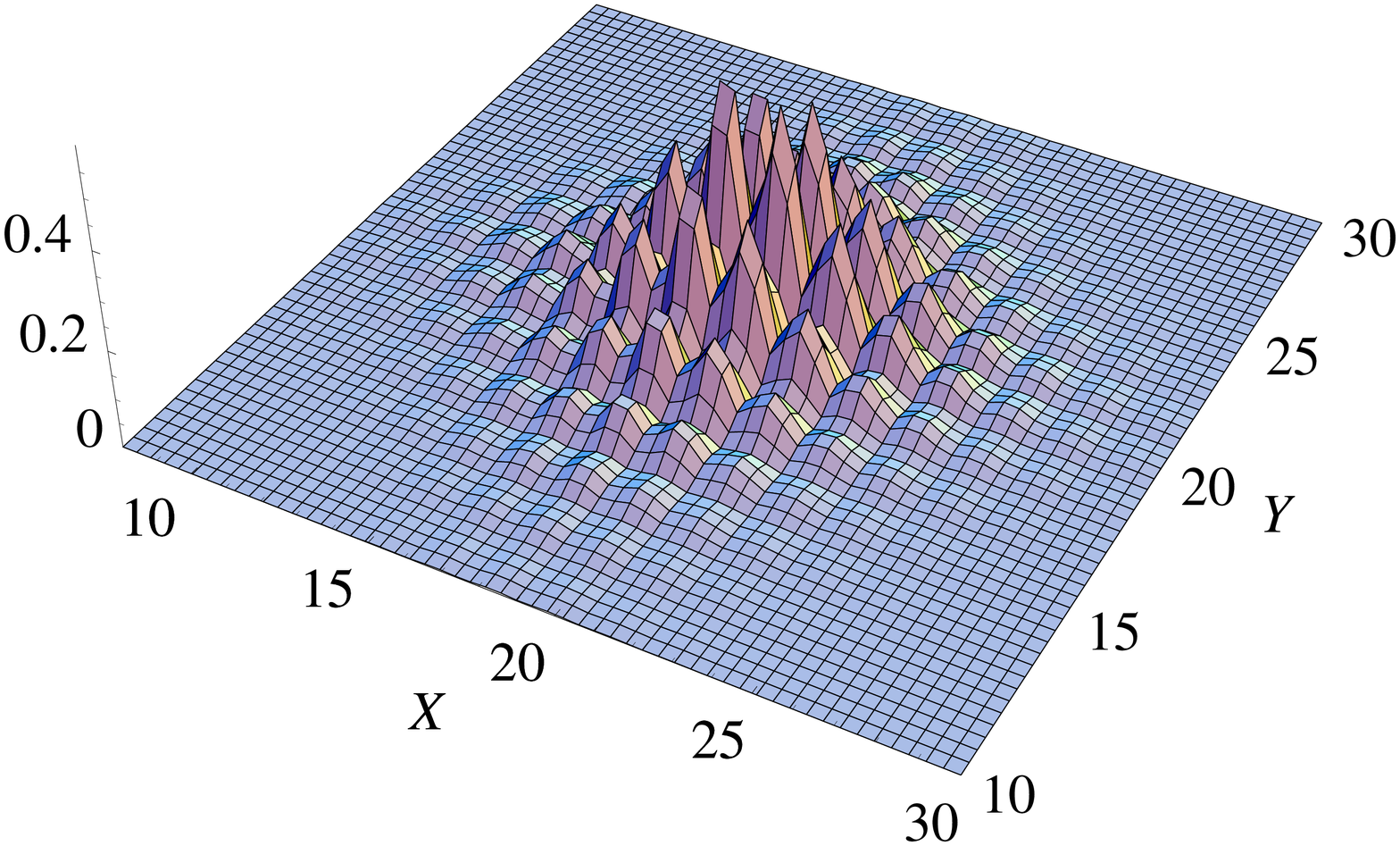}\caption{Detail of
the $q_{22}^{2}$ dromion interaction shown in figure
$4$.}\end{center}\end{figure}\\

\section{Conclusions}In this paper we have derived a noncommutative
version of the Davey-Stewartson equations and verified their
quasiwronskian and quasigrammian solutions by direct substitution.
The quasigrammian solution has then been used to obtain dromion
solutions in the matrix case, which, if we were to consider these
solutions in the scalar case, agree with the results of Gilson and
Nimmo in \cite{GN2}.  We have obtained plots of the $(1,1)$-dromion
solution and its plane waves, choosing the entries of the Hermitian
matrix $H$ in such a way that our solution is always well-defined.
In the $(2,2)$-dromion case, some of the more straightforward
asymptotic calculations have been carried out, enabling us to obtain
plots of the situation with one dromion vanishing as
$t\rightarrow{-\infty}$. \section*{Acknowledgements} Susan
Macfarlane would like to thank the Engineering and Physical Sciences
Research Council for a Research Studentship, and the reviewers of
this paper for helpful comments.
\appendix
\setcounter{section}{1}
\section*{Appendix}
Here we prove the results of section \ref{sec:derivqd}.  Consider a
general quasideterminant of the form given by (\ref{eq:Xi}).  Using
the product rule for derivatives,
\begin{equation}\label{eq:lambdadash}\Xi'=D'-C'A^{-1}B+CA^{-1}A'A^{-1}B-CA^{-1}B'.\end{equation}
We modify slightly the approach of \cite{GN} and find that, if $A$
is a Grammian-like matrix with derivative
\begin{equation}A'=\sum_{i=1}^{k}E_{i}F_{i},\end{equation} where
$E_{i}$ $(F_{i})$ are column (row) vectors of comparable lengths,
then
\begin{align}\notag\Xi'&=D'-CA^{-1}B+\sum_{i=1}^{k}(CA^{-1}E_{i})(F_{i}A^{-1}B)-CA^{-1}B'\\&=
\label{eq:lambdadashgramApp}\begin{vmatrix}
A&B\\C'&\boxed{D'}\end{vmatrix}+\begin{vmatrix}
A&B'\\C&\boxed{0}\end{vmatrix}+\sum_{i=1}^{k}\begin{vmatrix}
A&E_{i}\\C&\boxed{0}\end{vmatrix}\begin{vmatrix}
A&B\\F_{i}&\boxed{0}\end{vmatrix}.\end{align}Note here that $`0$'
denotes the $2\times{2}$ matrix $\bigl( \begin{smallmatrix} 0&0\\
0&0
\end{smallmatrix} \bigr)$.  If, on the other
hand, the matrix $A$ does not have a Grammian-like structure, we can
once again write the derivative $\Xi'$ as a product of
quasideterminants as above by inserting the $2n\times{2n}$ identity
matrix in the form
\begin{equation}I=\sum_{k=0}^{n-1}(f_{k}\,e_{k})(f_{k}\,e_{k})^{T},\end{equation}with
$e_{k}$, $f_{k}$ defined as before, so that $(f_{k}\,e_{k})$ denotes
the $2n\times{2}$ matrix with the $(2n-2k)^{\textrm{th}}$ and
$(2n-2k-1)^{\textrm{th}}$ entries equal to 1 and every other entry
0.  Then we find that
\begin{equation}\label{eq:lambdadashwronApp}\Xi'=\begin{vmatrix}A&B\\C'&\boxed{D'}\end{vmatrix}+\sum_{k=0}^{n-1}
\begin{vmatrix}A&\begin{array}{cc}f_{k}&e_{k}\end{array}\\C&\boxed{\begin{array}{cc}0&0\\0&0\end{array}}\end{vmatrix}
.\begin{vmatrix}A&B\\\begin{array}{c}(A^{2n-2k-1})'\\(A^{2n-2k})'\end{array}&
\boxed{\begin{array}{c}(B^{2n-2k-1})'\\(B^{2n-2k})'\end{array}}\end{vmatrix},
\end{equation}where $A^{k}$ denotes the $k^{\textrm{th}}$ row of $A$.
It is also possible to obtain a column version of the derivative
formula by inserting the identity in a different position.  We now
use the formulae (\ref{eq:lambdadashgramApp}) and
(\ref{eq:lambdadashwronApp}) to derive expressions for the
derivatives of the quasideterminants $Q(i,j)$, $R(i,j)$.\\
\
\\
\ Consider the quasiwronskian $Q(i,j)$ defined in section
\ref{subsec:qwsoln}, namely
\begin{equation}\label{eq:QijApp}Q(i,j)=\begin{vmatrix}\widehat{\Theta}&\begin{array}{cc}f_{j}&e_{j}\end{array}\\\Theta^{(n+i)}&
\boxed{\begin{array}{cc}0&0\\0&0\end{array}}\end{vmatrix}.\end{equation}Calculation
of the derivatives of $Q(i,j)$ requires knowledge of the following
result \cite{GN}, that for arbitrarily large
$n$,\begin{equation}\label{eq:cases}Q(i,j)=\begin{cases}-I_2&i+j+1=0,\\\bigl(
\begin{smallmatrix} 0&0\\ 0&0
\end{smallmatrix} \bigr)&\text{$i<0$
or $j<0$ and $i+j+1\neq{0}$}.\end{cases}\end{equation}We utilise the
dispersion relations for the ncDS system (\ref{eq:ds1}-d), found by
considering the Lax pairs (\ref{eq:L},b) in the trivial vacuum case,
giving, for $\theta$ an eigenfunction of $L$, $M$,
\begin{subequations}\begin{align}\label{eq:disp1}\theta_{x}&=-\sigma{J}\theta_{y},\\
\label{eq:disp2}\theta_{t}&=\textrm{i}J\theta_{yy},\end{align}\end{subequations}and
since $\Theta=\left(\theta_{1},\ldots,\theta_{n}\right),$ it follows
that \begin{subequations}\label{eq:Thetadisp}\begin{align}\label{eq:disp111}\Theta_{x}&=-\sigma{J}\Theta_{y},\\
\label{eq:disp211}\Theta_{t}&=\textrm{i}J\Theta_{yy}.\end{align}\end{subequations}Thus,
using (\ref{eq:lambdadashwronApp}), we have \begin{subequations}\begin{align}{Q}(i,j)_{y}&=Q(i+1,j)+\sum_{k=0}^{n-1}Q(i,k)Q(-k,j),\\
{Q}(i,j)_{x}&=-\sigma\left({J}Q(i+1,j)+\sum_{k=0}^{n-1}Q(i,k)JQ(-k,j)\right),\\
{Q}(i,j)_{t}&=\textrm{i}\left(JQ(i+2,j)+\sum_{k=0}^{n-1}Q(i,k)JQ(-k,j)\right).\end{align}\end{subequations}These
can be simplified using (\ref{eq:cases}), leaving the derivatives as
given in (\ref{eq:Qyxt}).\\
\
\\
\ We can apply a similar procedure to determine the derivatives of
the quasigrammian $R(i,j)$ defined in section \ref{subsec:qgsoln}.
The dispersion relations are found by considering the adjoint Lax
pairs (\ref{eq:Ladj},b) in the trivial vacuum case, giving, for
$\rho$ an eigenfunction of $L^{\dag}$,
$M^{\dag}$,\begin{subequations}\begin{align}\rho_{x}&=-\frac{1}{\sigma}{J}\rho_{y},\\\rho_{t}&=\textrm{i}J\rho_{yy},\end{align}\end{subequations}and
since $P=(\rho_{1},\ldots,\rho_{n})$, we have
\begin{subequations}\label{eq:Pdisp}\begin{align}P_{x}&=-\frac{1}{\sigma}{J}P_{y},\\P_{t}&=\textrm{i}JP_{yy}.\end{align}\end{subequations}
We also recall that from our construction of the binary Darboux
transformation in section \ref{subsec:BDT}, the potential
$\Omega(\phi,\psi)$ satisfies the relations (\ref{eq:omegay}-c),
from which it follows that
\begin{subequations}\begin{align}\Omega(\Theta,P)_{y}&=P^{\dag}\Theta,\\
\Omega(\Theta,P)_{x}&=-\sigma{P}^{\dag}J\Theta,\\\Omega(\Theta,P)_{t}&=\textrm{i}\left(P^{\dag}J\Theta^{(1)}
-P^{\dag(1)}J\Theta\right),\end{align}\end{subequations}where
$^{(k)}$ denotes the $k^{\textrm{th}}$ $y$-derivative.  Thus, using
(\ref{eq:lambdadashgramApp}), we calculate the derivatives of
$R(i,j)$ and find that they are identical to those of $Q(i,j)$ in
(\ref{eq:Qyxt}).

\bibliographystyle{plain}
\bibliography{GilsonMacfarlane18thMay}
\end{document}